\documentclass[aps,prb,twocolumn,floatfix]{revtex4}
\usepackage{graphicx}
\usepackage{amssymb}
\usepackage{amsmath}
\usepackage{bm}
\usepackage{epsf}

\newcommand{\bra}[1]{\langle #1|}
\newcommand{\ket}[1]{|#1\rangle}

\renewcommand{\phi}{\varphi}
\renewcommand{\epsilon}{\varepsilon}
\newcommand{\eff}{\textrm{eff}}
\renewcommand{\vec}[1]{{\bf #1}}

\begin{document}
\title{Topological characterization of periodically-driven quantum systems}
\author{Takuya Kitagawa}
\affiliation{Physics Department, Harvard University, Cambridge,
MA 02138, USA}
\author{Erez Berg}
\affiliation{Physics Department, Harvard University, Cambridge,
MA 02138, USA}
\author{Mark Rudner}
\affiliation{Physics Department, Harvard University, Cambridge,
MA 02138, USA}
\author{Eugene Demler}
\affiliation{Physics Department, Harvard University, Cambridge,
MA 02138, USA}

\begin{abstract}
Topological properties of physical systems can lead to robust
behaviors that are insensitive to microscopic details. Such
topologically robust phenomena are not limited to static systems
but can also appear in driven quantum systems.
In this paper, we show that the Floquet operators of periodically driven systems
can be divided into topologically distinct (homotopy) classes, and
give a simple physical interpretation of this classification in terms of
the spectra of Floquet operators.
Using this picture, we provide an intuitive understanding of the well-known
phenomenon of quantized adiabatic pumping.
Systems whose Floquet operators belong to the trivial class
simulate the dynamics generated by time-independent Hamiltonians,
which can be topologically classified according to the schemes developed for
static systems.
We demonstrate these principles through an example of
a periodically driven two--dimensional hexagonal lattice model
which exhibits several topological phases.
Remarkably, one of these phases supports chiral edge modes even though
the bulk is topologically trivial.

\end{abstract}

\date{\today}
\maketitle
%

\section{Introduction}
Following the discovery of the quantized Hall effect in 1980\cite{integerexperiment}, there has been great excitement about the possibility of observing extremely robust, ``topologically protected'' quantum phenomena in solid state systems
\cite{halperin,shinsei}.
This excitement has been redoubled in recent years with the discovery of new classes of materials called ``topological insulators'' in two and three dimensional systems\cite{fu1,bernevig,konig,Hsieh}.
A common feature linking all of these systems, which from a traditional point of view appear to be mundane band insulators, is the fact that their ground state wave functions feature internal structures characterized by non-zero values of integer topological invariants, which distinguish them from conventional, trivial systems.
The existence of such internal structures leads to the appearance of robust gapless edge modes wherever a non-trivial material has an interface with a trivial one, such as the vacuum.
Due to their topological origin, these modes can not be localized or destroyed by a wide range of perturbations, and give rise to many interesting, robust phenomena.
Recently, several groups have proposed a comprehensive scheme for classifying all possible types of such topological phases which can arise in band insulating and superconducting systems\cite{kitaev,schnyder,qi}.

Meanwhile, a variety of robust topologically-protected phenomena have also been found to occur in the dynamics of driven quantum systems with time-dependent Hamiltonians.
Such phenomena can roughly be divided into two broad classes.
First, there is a class of phenomena displaying quantized adiabatic pumping\cite{fu2,berg,aharony}, which can be conceptually traced to Thouless' original proposal of quantized adiabatic transport\cite{thouless,niu}.
Second, many groups have studied the possibility of effectively simulating
the behavior of the topologically non-trivial (static) materials described above by applying periodic driving fields to artificial
\cite{kitagawa,jaksch,sorensen, zhu, dalibard,spielman,lewenstein,fleishauwer} or condensed matter systems\cite{oka,tanaka,Lindner}.
In this paper we provide a unifying scheme for characterizing these various types of topological phenomena which occur in the dynamics of periodically driven systems.
Furthermore, from this unified view, we find that these two classes of phenomena are not unrelated, and can in fact both be realized
within a single periodically-driven graphene-like system.

Unlike the case of static (non-driven) systems,
periodically-driven systems do not have well-defined ground states which can be used for classification.
Instead, we classify driven systems in terms of the topological properties of their corresponding ``Floquet operators,'' i.e. their time evolution operators acting over one full period of the drive, $T$.
Each eigenstate of the Floquet operator, called a Floquet state, accumulates a phase $\phi$ over one period of the driving.
Accordingly, to each Floquet state, we associate a ``quasi-energy'' $\epsilon = \phi/T$, which is the average phase accumulated per unit time.
For many purposes, the Floquet states and their associated quasi-energies
can be regarded in an analogous way to the eigenstates and corresponding energies of a static system.
However, because the quasi-energy is defined as a phase variable, it is periodic with period $2\pi/T$.
This periodicity introduces an additional topological structure, associated with the winding of quasi-energy, 
which has no analogue in static systems.
As we will discuss below, this property allows a number of interesting phenomena to occur in driven systems, such as quantized pumping, and 
even the existence of chiral edge modes in a two-dimensional system with topologically-trivial bulk bands ({\it i.e.} with all Chern numbers equal to zero).

The mathematical structure which captures these quasi-energy
winding-related phenomena is that of the homotopy groups of unitary (Floquet) operators.
Throughout the main text, we will focus primarily on systems with discrete (lattice)
translational symmetry in $d$-dimensions, which are subjected to spatially homogenous, periodic, time-dependent driving.
In this case, we obtain simple expressions for topological invariants associated with the first and third homotopy groups, written in terms of Floquet operators parameterized by the conserved crystal momentum $\vec{k}$.
The case of disordered systems is discussed in terms of a ``twisted boundary condition'' approach\cite{niu}
in Appendix \ref{generalizednu1}.

When the homotopy group classification returns a trivial result, i.e. for systems without quasi-energy winding, the Floquet operator can be expressed in terms of a {\it local} effective Hamiltonian $H_{\rm eff}$ through $U(T) = e^{-i H_{\rm eff}T}$.
Here one can view the dynamics as a stroboscopic simulation of the dynamics of a static system governed by the Hamiltonian $H_{\rm eff}$.
In this case, the topological characteristics of the driven system can be classified in terms of the symmetry and dimensionality according to the schemes laid out for static topological insulators and superconductors in Refs.\onlinecite{kitaev,schnyder, qi}.
Just as in the static case, systems characterized by non-zero values of the topological invariants in these schemes possess robust chiral edge modes at interfaces with ``trivial'' regions\cite{halperin, shinsei}.

The characterization of periodically driven systems in terms of the topological structure of Floquet operators constitutes the major result of this paper.
This approach provides a natural description of topologically-quantized pumping, and reveals a simple and intuitive picture in which to understand this phenomenon.
Furthermore, the general mathematical structure provides a guide for identifying new classes of topologically-protected behavior which may arise in driven systems.
The intuition gained from this approach may help open new avenues in which to explore topologically-robust behaviors in artificial systems as well as in natural materials, where experimental realizations are already feasible with current technology.

The paper is structured as follows.
In Section \ref{setup}, we define the class of driven systems that we will consider, and review the basic
language of Floquet theory.
Then in Section \ref{evolutiontopology} we present the classification of driven systems in terms of the homotopy groups of their associated Floquet operators.
Here we show that the topological invariant $\nu_1$ associated with the first homotopy group of the Floquet operator directly measures the winding of the quasi-energy spectrum, and describe how it leads to the quantized average displacement.
In Section \ref{effectivehamiltonian} we describe the second scheme of topological classification of driven systems based on ``effective Hamiltonians,'' which applies in the absence of quasi-energy winding. Here we provide a detailed discussion of the meaning of effective time reversal and particle-hole symmetries for driven systems.
In Section \ref{graphene} we demonstrate the appearance of
both topological structures described in Section \ref{evolutiontopology}
and \ref{effectivehamiltonian}
in the dynamics within a two dimensional tight binding model on a hexagonal lattice with time dependent hopping amplitudes.
Strikingly, over a range of parameters, this model exhibits a phase which features chiral edge modes
in the absence of  topologically non-trivial bulk quasi-energy bands.
We conclude in Section \ref{conclusion}.

\section{Floquet theory  framework} \label{setup} In this section, 
we describe the Floquet theory framework that will be used to analyze the dynamics of driven systems in this paper. 
We focus on the driven dynamics of a quantum system of non-interacting particles in a
$d$-dimensional lattice, subjected to a periodically-varying,
time-dependent Hamiltonian $H(t)$, which satisfies $H(t + T) = H(t)$.
Here $T$ is the period of the driving cycle.
The behavior is analyzed in terms of the evolution operator of the system over one full period
of the driving, $U(T)$, 
defined as
\begin{equation}\label{ev}
U(T) = \mathcal{T} e^{-i \int^{T}_{0} H(t) dt},
\end{equation}
where $\mathcal{T}$ is the time-ordering operator.
In the theory of periodically-driven systems\cite{floquet}, $U(T)$ is called the Floquet operator.
Note that while $U(T)$ implicitly depends on the starting point of the interval, here taken to be $t = 0$, the topological classification described below is independent of the choice of the starting point.

\begin{figure}[t]
\begin{center}
\includegraphics[width = 7cm]{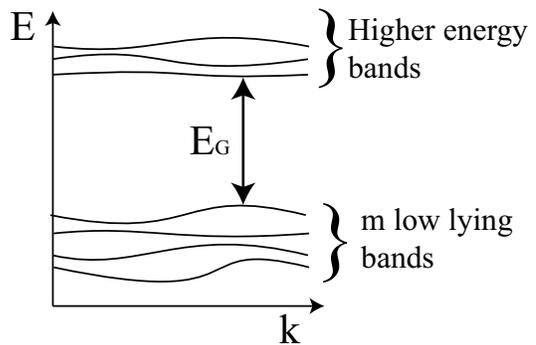}
\caption{
Typical schematic {\it instantaneous band structure} considered in this work, illustrated for a one-dimensional system with crystal momentum $k$.
We consider the case where $m$ low-lying bands are separated from higher bands by a band gap $E_G$ at all times $t$ during the periodic driving cycle.
We assume that the frequency of driving 
is much smaller than the band gap $E_{G}$ at each instant, such that excitations out of the $m$-fold low-lying band subspace can be neglected.}  
\label{bandgap}
\end{center}
\end{figure}
Throughout the main text, we assume that the system possesses discrete (lattice) translational symmetry at each
instant of time.
However, topological classification can still be applied even when the system lacks translational invariance.
The extension to this more general case is discussed in Appendix \ref{generalizednu1}.
In the presence of translational invariance, the crystal momentum $\vec{k}$ is a conserved quantity of the dynamics.
Therefore, at each 
time $t$, the Hamiltonian $H(t)$ can be characterized by a band structure as shown schematically in Fig.\ref{bandgap}.

Suppose that at time $t=0$, the state of the system lies entirely
within the subspace spanned by the $m$ lowest bands of the
instantaneous Hamiltonian $H(0)$. In this work we consider the
situation where the driving always returns the system to this
subspace after each full period, without causing excitations to
other bands.
This condition is guaranteed, for example, when higher energy bands of $H(t)$ are well separated
from lower energy bands by a large energy gap $E_g$(see Fig.\ref{bandgap}) and the evolution is adiabatic with respect to $E_g$ at all times.
In this situation, the evolution operator 
for one full period of driving is described by the set of $m\times
m$ matrices $\{U_{\vec{k}}(T)\}$, labelled by the value of the
crystal momentum $\vec{k}$. In terms of these operators, the
Floquet operator, Eq.(\ref{ev}), projected onto the low-energy
subspace, is given by $U(T) = \sum_{\vec{k}} U_{\vec{k}}(T)
\otimes P_{\vec{k}}$, where $P_{\vec{k}}$ is the projector onto
the $m$-dimensional subspace of low-energy states with crystal
momentum $\vec{k}$. Note that for intermediate times $0 < t < T$,
even in the fully adiabatic regime, the evolution operator
$U_{\vec k}(t)$ will in general include mixing of higher bands of
the initial Hamiltonian $H(0)$ and therefore {\it cannot} be represented in the basis of
the $m$ lowest bands of $H(0)$. The restriction to $m$ dimensions for $H(0)$
applies {\it only for complete cycles} of the periodic driving.


In analogy with the energy associated with each eigenstate of the Hamiltonian of
a time-independent system, we associate a quasi-energy $\varepsilon$ with
each eigenstate $\ket{\phi}$ of the Floquet operator of the driven system.
The quasi-energy is defined through the phase accumulated by the ``Floquet state''
$\ket{\phi}$ over each full period of driving as $U(T)\ket{\phi}=e^{-i\varepsilon T} \ket{\phi}$.
Note that, unlike real energies which are uniquely determined, quasi-energies are only uniquely defined up to integer multiples of $2\pi/T$.
This property will be crucial to the analysis below.
For systems with discrete translational symmetry, we index the quasi-energies by the crystal momentum $\mathbf{k}$, and define a ``quasi-energy band structure'' $\{\varepsilon_{\mathbf{k},\alpha}\}$, where $\alpha$ is a band index.
In the next section, we will see that the ``winding'' property of quasi-energy allows for quasi-energy band structures, and hence behaviors, which are qualitatively different from those found in static systems.
\section{Homotopy group classification of evolution operators} 
\label{evolutiontopology} In this section, we formulate the
topological classification of periodically driven systems in terms
of the homotopy groups of the evolution operators
$\{U_{\vec{k}}(T)\}$. In subsection \ref{1dinvariant}, we focus on
the topological invariant $\nu_1$ associated with the first homotopy group
of evolution operators. We illustrate many general
properties of the topology in evolution operators through the
study of an example in one dimension. Then in subsection
\ref{3dinvariant} we discuss generalizations to higher dimensions.

\subsection{Topological invariant $\nu_1$ and quasi-energy winding}
\label{1dinvariant} 
In one spatial dimension, the set of operators $\{U_k(T)\}$ defines a map from the Brillouin zone $-\pi \le k < \pi$ to the space of $m\times m$
unitary matrices.
Due to the periodicity of crystal momentum, the Brillouin zone is a circle, and this map traces out a closed loop in the space of $m \times m$ unitary matrices.
This loop can be characterized in terms of a homotopy class, which identifies all such loops that can be smoothly deformed into one another.
These homotopy classes are indexed by an integer-valued topological invariant $\nu_1$ that captures the topology (or ``winding'') of the corresponding maps.

In the presence of translational symmetry, the invariant $\nu_1$ is defined as:
\begin{equation}
\nu_{1}  = \frac{1}{2\pi} \int_{-\pi}^\pi dk\, \textrm{Tr} \Big[
U_k(T)^{-1} i\partial_{k} U_{k}(T) \Big],  \label{nu1}
\end{equation}
where $k$ is integrated over the first Brillouin zone, and the
trace is taken over the $m$-dimensional internal space of the Bloch wave functions\cite{footnote_basis}.
We emphasize that a map characterized by the topological invariant $\nu_1$ can be defined in a system with spatial dimension $d > 1$ by choosing $k$ in Eq.(\ref{nu1}) to parametrize some closed, non-contractible loop in the $d$-dimensional Brillouin zone.
For example, consider the case of a two-dimensional system on a square lattice with 
crystal-momentum components $k_{x}$ and $k_{y}$.
Here the Brillouin zone is a torus, and we can construct two invariants $\nu_{1x}$ and $\nu_{1y}$ associated with 
two distinct loops on the torus, $k_x = {\rm constant}, -\pi \le k_y < \pi$, and $-\pi \le k_x < \pi, k_y = {\rm constant}$:
$\nu_{1x} = \frac{1}{2\pi} \int_{-\pi}^\pi dk_{y} \, \textrm{Tr} \Big[
U_{\vec k} (T)^{-1} i\partial_{k_{y}} U_{\vec k}(T) \Big]$ for a fixed $k_{x}$, and similarly for $\nu_{1y}$. 
Note that, due to continuity, $\nu_{1x}$ ($\nu_{1y}$) is independent of the value of $k_x$ ($k_y$).
Hence this first homotopy group classification is not restricted to systems in one spatial dimension.


\begin{figure}[t]
\begin{center}
\includegraphics[width = 8.5cm]{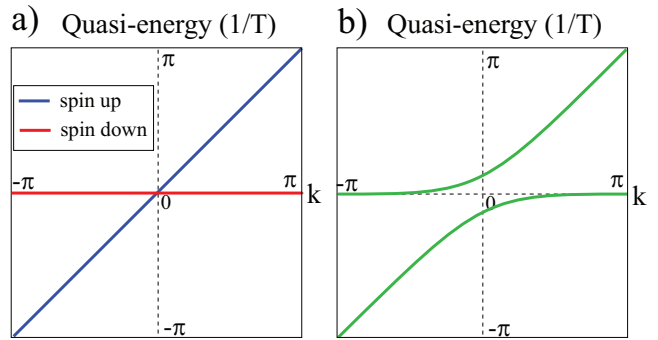}
\caption{a) Quasi-energy spectrum of a spin-$1/2$ particle in a time- and spin-dependent potential, displaying non-trivial topology characterized by $\nu_{1} =1$ (see text).
The quasi-energy is plotted in units of $1/T$.
The topological invariant $\nu_{1}$ counts the total winding of the bands in the quasi-energy direction as crystal momentum $k$ is taken around the Brillouin zone.
b) The total winding number $\nu_{1}$ is unchanged when coupling between up and down spin states is introduced. }
\label{1ddynamicaltopologymix}
\end{center}
\end{figure}
The meaning of the topological invariant $\nu_1$ is best
illustrated through a simple example. Consider a spin-$1/2$
particle in a deep one-dimensional, spin-dependent lattice potential.
 Here and throughout the paper, we take the lattice constant to be $1$.
The potential for the spin-down state is time-independent, while
the potential for the spin-up state moves slowly to the
right, such that over the time $T$ it shifts by exactly $s$ times the lattice constant, where $s$ is an integer.
If we assume that the movement is adiabatic, 
then the Floquet operator restricted to the space of the lowest band of initial Hamiltonian $H(0)$ is given by
$U(T) = \sum_{x} \ket{x+s}\bra{x} \otimes P_{\uparrow} + 1 \otimes P_{\downarrow}$, where $P_{\sigma}=\ket{\sigma}\bra{\sigma}$ is a projector onto the spin state $\sigma=\uparrow,\downarrow$, and $\ket{x}$ is a state localized in a single well of the (deep) lattice potential, with $x$ an integer labeling the unit cell.
In the crystal momentum basis, we have 
$U_k(T) = e^{-isk} \otimes P_{\uparrow} +1 \otimes P_{\downarrow}$.
This evolution is indeed characterized by the non-trivial value $\nu_1 = s$, as can be checked
explicitly using Eq.(\ref{nu1}).

We can obtain an intuitive understanding of $\nu_1$ by evaluating Eq.(\ref{nu1}) in the basis of Floquet states.
This analysis gives
\begin{equation}
\nu_{1} = \sum_{\alpha} \frac{1}{2\pi} \int^{\pi}_{-\pi}
dk\,\frac{d\epsilon_{k,\alpha}}{dk}T, \label{winding}
\end{equation}
where $\{\epsilon_{k,\alpha}\}$ are the quasi-energy bands labeled by the band index
$\alpha$\cite{degeneracy}. The integral in Eq. (\ref{winding})
counts the total winding number of the quasi-energy bands as $k$
runs over the first Brillouin zone. This picture of $\nu_1$ is
particularly appealing since it allows the non-trivial topology to
be identified through simple inspection of the quasi-energy
spectrum. The quasi-energy spectrum for the case $s=1$ is plotted
in Fig.\ref{1ddynamicaltopologymix}a. Note that such winding is only possible due to the periodicity of quasi-energy for a driven
system; therefore $\nu_1=0$ for any local, static (non-driven) system.

An important physical manifestation of the topology captured by $\nu_1$ is revealed by Eq.(\ref{winding}).
The factor $d\epsilon_{k,\alpha}/dk$ plays the
role of a group velocity for the Floquet band $\alpha$.
Due to the periodicity of $\epsilon_{k,\alpha}$ in $k$ and the periodic nature of quasi-energy, the average group velocity is
quantized: the average slope of $\epsilon_{k,\alpha}$ must take a value which yields an integer number of windings over the Brillouin zone.
Averaged over all $k$, the displacement $\Delta x =
\overline{({d\epsilon_{k,\alpha}}/{dk})}\cdot T$ after a full
period $T$ is therefore quantized. In the simple example above, in
which the particle is initially in a uniform superposition of
states with all values of $k$, this quantization means that over
one period the average position of a particle in the spin-up band
is shifted by {\it exactly} $s$ unit cells to the right, while the
average position of a particle in the spin-down band does not
change.
For the case of  adiabatic evolution of a filled-band fermionic system,
it can be shown that $\nu_1$ is directly related to the charge current integrated over one period. Therefore,
in this case, the quantization of $\nu_1$ implies the quantization of pumped charge
first identified by Thouless\cite{thouless}. In Appendix \ref{pump}, we explicitly
show this relation. Moreover, as is implied by the relation between $\nu_1$ and quantized charge pumping, $\nu_1$ can be directly related to the first Chern number of the Bloch
wavefunctions of the filled bands. We demonstrate the general relation between the topology captured by homotopy groups of Floquet operators and Chern numbers in
Appendix \ref{chernnumbers}.

While the homotopy-class-based topological characterization of periodically driven systems given in this paper is closely related to the Chern-class-based characterization of adiabatic pumping developed in previous works\cite{thouless,qi}, we emphasize that our approach provides an intuitive framework which naturally leads the way to generalizations to systems in other dimensions or with additional symmetries (see Section \ref{3dinvariant}).
  Moreover, as illustrated by an example in Section \ref{graphene}, the picture based on the winding of quasi-energy makes it possible to identify non-trivial topological behavior of driven systems in a simple, direct manner.

The integer-quantization of $\nu_1$ implies that its value is robust against
various perturbations. For instance, the value of $\nu_1$ is insensitive to continuous deformations of the quasi-energy spectrum,
and to mixing with other, topologically trivial, quasi-energy bands.
For the model described above with quasi-energy band structure shown in
Fig.\ref{1ddynamicaltopologymix}a, such mixing can be introduced,
for example, by applying a spin rotation pulse during the dynamics.
The quasi-energy spectrum in the presence of such mixing is shown schematically in
Fig.\ref{1ddynamicaltopologymix}b.
Note that, even in this case, Eq.(\ref{winding}) for the net winding number still gives $\nu_1 = 1$;
hybridization of Floquet bands with trivial and non-trivial topology does not change the total winding number.


In the presence of disorder, translational invariance is
destroyed. The crystal momentum $k$ is not a good quantum number, and the definition of $\nu_1$ in  Eq.(\ref{nu1}) cannot be used.
Nonetheless, the topologically protected phenomena are expected to be robust against weak perturbations, including disorder.
Indeed, in Appendix \ref{generalizednu1}, we use the method of twisted boundary
conditions\cite{niu} to define a generalization of the topological
invariant $\nu_1$, which is valid even in the presence of
disorder. This generalized invariant can also be used to classify the
dynamics of periodically driven, gapped, interacting many-body
systems, such as filled-band states of fermions or Mott insulating
states of bosons {\cite{berg}}.

\subsection{Higher dimensional systems} \label{3dinvariant}
In the previous subsection, we saw that the phenomenon of quantized adiabatic pumping in one dimension acquires a simple and intuitive explanation when analyzed in terms of the topological structure of Floquet operators.
We now discuss how this elegant mathematical framework can guide us in searching for generalizations of topological pumping in higher dimensional systems.
The key observation is that the Floquet operators of periodically driven systems with dimension greater than one can be characterized by topologies associated with homotopy groups beyond the fundamental group.
Here we describe the general situation in higher dimensions, and give an explicit expression for the topological invariant associated with the third homotopy group, which can be relevant in three-dimensional systems.

For a system in $d$ dimensions, the first Brillouin zone is topologically equivalent to a $d$-dimensional torus.
By allowing $\vec{k}$ to parametrize an $\ell$-dimensional toroidal section of the Brillouin zone, with $\ell \le d$, we can use the set of (Floquet) evolution operators $\{U_{\vec{k}}(T)\}$ to define a map from the $\ell$-dimensional torus to the space of $m \times m$ unitary matrices, ${\rm U}(m)$.
Such maps can be classified using the structure of the $\ell$-th homotopy group of ${\rm U}(m)$, denoted by $\pi_{\ell}[{\rm U}(m)]$.
Noting that the homotopy groups are given by $\pi_{d}[{\rm U}(m)]=\mathbb{Z}$ for odd $d$ and $\pi_{d}[{\rm U}(m)]=1$ for even $d$, with $m \ge (d + 1)/2$\cite{nakahara}, we see that a new type of topological invariant appears in each {\it odd} dimension.
While the $d$-th homotopy group of generic unitary matrices is trivial when $d$ is even, the Floquet operators of even-dimensional systems can still possess non-trivial topological structures either under lower homotopy groups (e.g. $\pi_1[{\rm U}(m)]$ for a system in $d=2$), or if the space of allowed evolution operators is restricted by additional symmetries of the system.

For systems in three dimensions, $d = 3$, we consider the topological invariant $\nu_3$ associated with the third homotopy group:
\begin{eqnarray}
\nu_{3} & = & \int  \frac{d^3 k}{24 \pi^2}\ \epsilon^{\alpha \beta \gamma}\times \\ &&\textrm{Tr}\Big[  \left(U^{-1}_{\vec{k}}\partial_{k_\alpha}U_{\vec{k}}\right) 
\left(U^{-1}_{\vec{k}}\partial_{k_\beta}U_{\vec{k}}\right) \left(U^{-1}_{\vec{k}}\partial_{k_\gamma}U_{\vec{k}}\right) \Big],\nonumber
\label{nu3}
\end{eqnarray}
where 
$\epsilon^{\alpha \beta \gamma} $ is
the anti-symmetric tensor, $\alpha, \beta, \gamma = {x}, {y}, {z}$, and $\vec{k}$ is integrated over the first Brillouin zone.
Note that $\nu_3$ can only be non-vanishing if the dimension of $U_{\vec{k}}(T)$ is larger than $1$ so that multiplication can be non-commutative. The search for physical realizations and the manifestations of this topological
invariant will be interesting subjects for future work.


The phenomena associated with non-trivial topology of Floquet operators under the homotopy groups are unique to periodically-driven systems.
For a system governed by a static Hamiltonian $H_S$, the eigenstates of the Floquet operator $U_S(T)= e^{-iH_ST}$ and the associated quasi-energies for any choice of $T$ coincide with those of $H_S$.
Because the energy eigenvalues $E_{\vec{k},n}$ of any local Hamiltonian $H_S$ are smooth, periodic functions of $\vec{k}$, the corresponding quasi-energy eigenvalues $\epsilon_{\vec{k},n}$ {\it cannot} wind and there must exist a gap in the quasi-energy spectrum.
It can be shown that 
a ``gapless'' quasi-energy spectrum, in which there is at least one eigenstate for every quasi-energy $-\pi/T \leq \epsilon < \pi/T$, is a necessary requirement for non-trivial topology under any homotopy group\cite{footnote_homotopy}. An example of such a gapless spectrum is shown in Fig. \ref{1ddynamicaltopologymix}. 
Therefore the Floquet operator of any local, static system is trivial under all homotopy groups.
For a periodically-driven system with driving period $T$, however, the quasi-energy spectrum associated with the evolution operator $U(T)$ need not have a gap.
Hence, as demonstrated by the example above, periodic driving {\it can} produce evolution operators with non-trivial topology under the homotopy groups.

\section{Topological invariants of gapped effective Hamiltonians}\label{effectivehamiltonian}
So far, we classified the dynamics of periodically driven systems
with discrete translational symmetry through the generalized
``windings'' of their Floquet evolution operators as identified by
the homotopy groups of unitary matrices. However, even when there
is a gap in a system's quasi-energy spectrum and its Floquet
operator is thus trivial under all homotopy groups, its dynamics
may possess other topological characteristics associated with an
effective Hamiltonian $H_{\rm eff}$:
\begin{equation}\label{heff}
U(T) = \mathcal{T}e^{-i \int^{T}_{0} H(t) dt} \equiv e^{-iH_{\textrm{eff}}T}.
\end{equation}
When viewed in this way, evolution under the periodically varying Hamiltonian $H(t)$ stroboscopically simulates the evolution of a system with a \emph{static} 
Hamiltonian $H_{\textrm{eff}}$ at integer multiples of the driving period $T$.

When $H_{\textrm{eff}}$ has a band structure with a band gap, the
previously introduced classification
schemes\cite{kitaev,schnyder, qi} for static insulators and
superconductors can be directly applied to $H_{\textrm{eff}}$\cite{kitagawa}.
In these schemes, systems are categorized into distinct topological classes based on dimensionality and on the presence or absence of the (discrete) time-reversal and particle-hole symmetries.
In a periodically-driven system, these symmetries must be considered at the level of the effective Hamiltonian, or equivalently, of the Floquet operator $U(T)$, as we now describe.

At the level of the Floquet operator $U(T)$, time-reversal symmetry is defined by the existence of a unitary operator $\mathcal{Q}$ which has the action
$\mathcal{Q} H_\mathrm{eff}^* \mathcal{Q}^\dagger =
H_{\mathrm{eff}}$, or
\begin{equation}
\mathcal{Q} U(T)^* \mathcal{Q}^\dagger = U(T)^\dagger.
\label{TR}
\end{equation}
Here $H_\mathrm{eff}^*$ is the complex conjugate of $H_\mathrm{eff}$.
The operator $\mathcal{Q}$ satisfies $\mathcal{Q} \mathcal{Q}^*=\pm 1$, where the sign, $+$ or $-$, distinguishes the cases with integer or odd-half-integer spin, respectively.  
A sufficient, although not necessary, condition for the presence of time-reversal symmetry is satisfied if the time-dependent Hamiltonian possess a special point $t_0$ such that
\begin{equation}\label{TR1}
\mathcal{Q} H(t+t_0)^* \mathcal{Q}^\dagger = H(-t+t_0).
\end{equation}
We give a detailed proof of this statement in Appendix \ref{sec:TR}.
Note that condition in Eq. (\ref{TR1}) can be satisfied even if the instantaneous Hamiltonian $H(t)$ is {\it not} time-reversal invariant, such as when the system is subjected to magnetic fields.
Conversely, even if $H(t)$ instantaneously has time-reversal symmetry for each $t$, the condition in Eq. (\ref{TR1}) is not necessarily satisfied, and $H_{\mathrm{eff}}$ need not be time reversal invariant.
This latter situation arises in the example which will be explored in Sec.\ref{graphene}.

Similarly, particle-hole symmetry is defined at the level of the Floquet operator by the existence of a local unitary operator $\mathcal{P}$ satisfying
\begin{equation}
\mathcal{P} U(T)^* \mathcal{P}^\dagger = U(T) \label{PH}
\end{equation}
with $\mathcal{P} \mathcal{P}^*=\pm 1$.
Any Bogoliubov-de Gennes Hamiltonian describing the dynamics of superconducting quasiparticles possesses this symmetry (with $\mathcal{P} \mathcal{P}^*= 1$), even if a time-dependent perturbation is added.


Many results from the study of topological properties in static systems can be directly
translated to periodically driven systems, with the understanding that the
topologically-protected phenomena apply to the behaviors of Floquet states. For example, the edge of a static
two dimensional system characterized by a non-zero Chern number is
known to host chiral edge states. In Section \ref{graphene}, we
study dynamics in a driven two-dimensional tight-binding system, 
and show that driving can induce non-zero Chern numbers in the bands of the effective Hamiltonian.
The edge of such a system hosts chiral Floquet edge states which propagate unidirectionally when viewed
at integer periods of the driving.
Analogously, the edge of a system with a time-reversal-invariant effective Hamiltonian can host helical Floquet edge states\cite{Lindner}.

\section{Dynamically-induced topological phases in a hexagonal lattice} \label{graphene}
In this section, we study single-particle dynamics in a two
dimensional hexagonal lattice tight-binding model, where the
hopping amplitudes are varied in a spatially uniform, but
time-dependent, cyclic fashion.
Here we choose the hexagonal lattice
but expect that similar physics can be realized in other lattices as well.
In different parameter regimes,
the system can support topological phases of either the Floquet operator homotopy type, or of the effective Hamiltonian type.
 For weak driving, there is no winding of the quasi-energy bands
($\nu_1 = 0$), but the driving induces non-zero first Chern
numbers in the two bands of the effective Hamiltonian. For larger
driving amplitudes, the Chern numbers associated with each of the
two bands become zero. While the bulk topology given by the Chern number is trivial,
the invariant $\nu_{1}$ becomes non-zero in a system with edges.

\begin{figure}
\begin{center}
\includegraphics[width = 8cm]{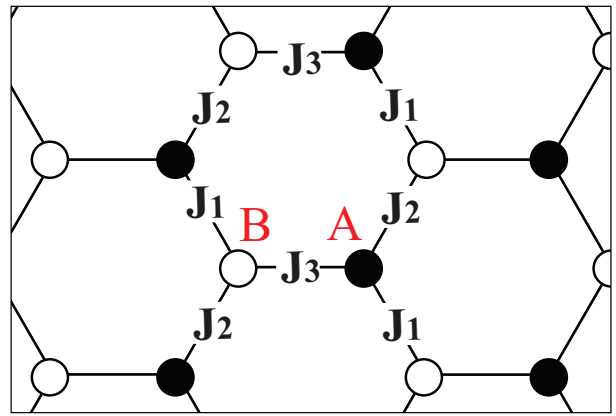}
\caption{Hexagonal lattice structure. Sublattice A is marked with
filled circles and sublattice B is marked with open circles.
$J_{i}$ for $i=1,2,3$ represent the hopping amplitudes between the
sites.} \label{lattice_fig}
\end{center}
\end{figure}
We start from a tight-binding model Hamiltonian on a hexagonal
lattice:
\begin{eqnarray}
H &=& \sum_{\vec{k}} \left( \begin{array}{cc} c^{\dagger}_{\vec{k}, A} & c^{\dagger}_{\vec{k},B} \end{array} \right)
 H(\vec{k})
 \left( \begin{array}{c} c_{\vec{k}, A} \\ c_{\vec{k},B} \end{array} \right)  \\
 H(\vec{k}) &=& -  \sum_{i} J_{i}(t) \left(\cos(\vec{b}_{i}\cdot \vec{k})  \sigma_{x}
+ \sin(\vec{b}_{i}\cdot \vec{k})  \sigma_{y}\right), \nonumber
\end{eqnarray}
where $A$ and $B$ label the two sublattices of the hexagonal lattice, 
$\{J_{i}\}$ are the hopping amplitudes from $B$ sites to the
neighboring $A$ sites in the three directions $i=1,2,3$ (see Fig.
\ref{lattice_fig}), and $\{\vec{b}_{i}\}$ are the vectors given by
$\vec{b}_{1} = (-1/2,\sqrt{3}/2)$, $\vec{b}_{2} =
(-1/2,-\sqrt{3}/2)$ and $\vec{b}_{3} = (1, 0)$. Here
$c^{\dagger}_{\vec{k}, \alpha}$ is the creation operator for a
Bloch state with crystal momentum $\vec{k}$ on sublattice $\alpha
= A, B$. Expressed in terms of local (site-specific) creation
operators $\{c^{\dagger}_{\vec{x}_{i},\alpha}\}$, we have
$c^{\dagger}_{\vec{k},\alpha} =
\frac{1}{\sqrt{N}}\sum_{\vec{x}_{i}}
c^{\dagger}_{\vec{x}_{i},\alpha} e^{i\vec{k} \cdot \vec{x}_{i}} $,
where $N$ is the number of unit cells in the system.
 Here and in the following, the distance between any site and its nearest neighboring sites in the hexagonal lattice is taken to be $1$.

\begin{figure}
\begin{center}
\includegraphics[width = 8.5cm]{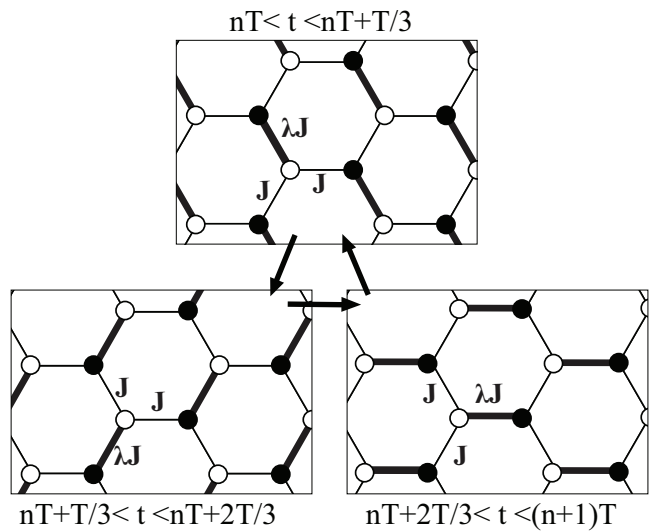}
\caption{Driving cycle considered in the text, in which the hopping amplitudes $\{J_i\}$ are varied in a cyclic fashion.
Here we consider only $\lambda \ge 1$, where the hopping amplitude along one of the three bond types is uniformly increased relative to the other two during each stage of the cycle. }
\label{hexagonallatticedynamics}
\end{center}
\end{figure}
We consider a driving protocol where the hopping amplitudes
$\{J_i(t)\}$ are modulated cyclically in time according to (see
Fig.\ref{hexagonallatticedynamics}):
\begin{enumerate}
\item $J_1 = \lambda J;\, J_2,J_3 = J$ for $nT< t \leq nT+ T/3$  
\item $J_2 = \lambda J;\, J_1,J_3 = J$ for $nT + T/3< t \leq nT+ 2T/3$ 
\item $J_3 = \lambda J;\, J_1,J_2 = J$ for $nT+ 2T/3< t \leq nT+  T$. 
\end{enumerate}
In the following, we consider the cases $\lambda \geq 1$. Notice
that when $\lambda = 1$, the Hamiltonian describes hopping on a
time-independent hexagonal lattice with uniform hopping
amplitudes. Just as in graphene, in this case the spectrum is
gapless, with two inequivalent Dirac points. We focus on the high
frequency limit, $J/\omega \ll 1$, where $\omega = 2\pi/T$ is the
angular frequency of the drive.  In this limit, all states within
the two bands are far from resonance, and the graphene-like band
structure provides a good zeroth-order approximation to the
quasi-energy spectrum.

We are interested in the topological properties of the dynamics
when $\lambda$ is increased from $1$. For $\lambda>1$, we obtain
the effective Hamiltonian through
\begin{equation}
e^{-iTH_{\eff}} \equiv e^{-iH_{3}T/3} e^{-iH_{2}T/3}e^{-iH_{1}T/3},
\end{equation}
where $H_{j}$ is the Hamiltonian during the time $nT+ (j-1)T/3< t \leq nT+ jT/3$ (see protocol above).
For $\lambda > 1$, gaps open at the Dirac points in the quasi-energy spectrum of $H_{\eff}$.
For small $(\lambda-1) \ll 1$, the size of the gap is given by $\sqrt{3} [(\lambda-1)J]^2 T$.
Because $H_{\rm eff}$ is a two-band Hamiltonian, we can
write\cite{footnoteeffectiveh}
\begin{equation}
H_{\eff}(\vec{k}) = \epsilon(\vec{k})\, \vec{n}(\vec{k}) \cdot {\bm \sigma},
\end{equation}
where ${\bm \sigma} = (\sigma_{x}, \sigma_{y}, \sigma_{z})$ is a vector of
Pauli matrices acting in the sublattice space, and $\vec{n}(\vec{k})$ is a three dimensional unit vector.
When $\epsilon(\vec{k}) \neq 0, \pi/T$ for all $\vec{k}$, each quasi-energy band of the
effective Hamiltonian can be characterized topologically by the
first Chern number. The first Chern numbers for the two bands are given by\cite{TKNN}:
\begin{equation}
\label{chern}
C_\pm = \frac{\pm 1}{4\pi} \int_{\rm FBZ}  \vec{n} \cdot (\partial_{k_{x}} \vec{n} \times \partial_{k_{y}} \vec{n} )d^2 \vec{k}.
\end{equation}
Here, the $\vec{k}$ integration is taken over the first Brillouin
zone.

We numerically calculate the Chern numbers $C_\pm$ from Eq.(\ref{chern}) 
using a fixed ratio $J/\omega= \frac{3}{32}$ between the hopping strength $J$ and the frequency $\omega$.
We find $C_\pm = \pm 1$ for $1<\lambda < \lambda_{c}$, where $\lambda_{c} \approx 3.3$ is the critical coupling where a topological phase transition occurs. 
At $\lambda = \lambda_{c}$, the quasi-energy gap at $\epsilon =
\pm \pi/T$ closes for the states at $k = 0$.
For $\lambda > \lambda_{c}$, the Chern numbers of both bands become zero until another phase transition
point $\lambda = \lambda'_{c}$, where $\lambda'_c \approx 8.7$ is
reached\cite{footnotechern}. Thus we see that the time-dependent driving within a
hexagonal lattice tight-binding model can induce 
non-zero Chern numbers in the effective Hamiltonian.

\begin{figure}
\begin{center}
\includegraphics[width = 8cm]{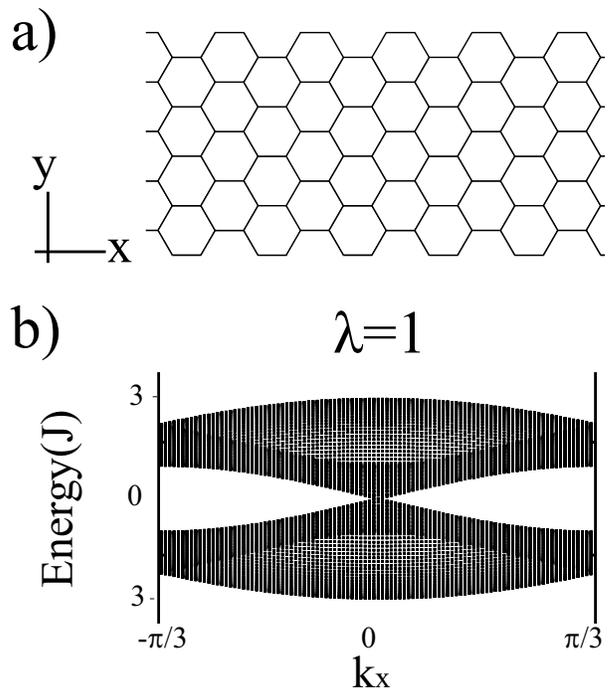}
\caption{a) The strip geometry considered in the text, with finite
extent in the $y$ direction, and armchair edges running parallel to $x$.
b) Spectrum of the (static) 
armchair ribbon shown in panel a) with a width of $100$ lattice sites in the $y$ direction. Energy is plotted
in units of $J$. 
} \label{zigzag}
\end{center}
\end{figure}

A consequence of non-zero Chern number in a system with boundaries
is the existence of chiral Floquet edge states. 
For example, a system with a ``strip'' geometry (see Fig.
\ref{zigzag}a) is expected to have two counter-propagating chiral
edge modes localized on the upper and lower boundaries.
In the following, we consider the strip geometry with edges of
armchair type.

\begin{figure}
\begin{center}
\includegraphics[width = 8.5cm]{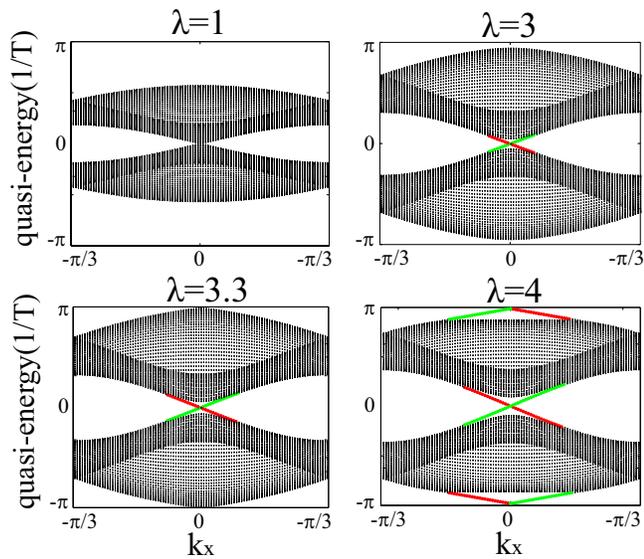}
\caption{Quasi-energy spectrum of $H_{\eff}$ in a finite strip of
width $100$ lattice sites in the $y$ direction with armchair edges
along the $x$ direction. Here we choose $JT =
\pi/16$, and $\lambda=1, 3, 3.3, 4$. For $\lambda =  3$, the two bands
of $H_{\eff}$ are characterized by non-zero Chern numbers, which are manifested in the presence
of chiral edge modes.
The lines corresponding to these modes are colored such that the mode localized on the upper (lower) edge is shown in red (green).
Close to $\lambda = 3.3$, the gap of $H_{\eff}$ at quasi-energy $\pi/T$ closes, and then opens again as $\lambda$ is increased further.
At $\lambda =4$, the Chern numbers associated with each of the bands are zero, yet the nanoribbon clearly still supports chiral edge states.}
\label{chiraledge}
\end{center}
\end{figure}

In Fig. \ref{chiraledge}, we plot the quasi-energy spectrum of the
driven strip as a function of the crystal momentum parallel to the strip, for several values of $\lambda$.
When $\lambda = 1$, the Hamiltonian is independent of time, and the quasi-energy
spectrum is identical to the true energy spectrum of a graphene ribbon with armchair edges.
For $\lambda= 3$, chiral edge states are
visible within the gap centered at quasi-energy $\epsilon = 0$. The states
are color-coded such that states which are localized at the upper
edge are shown in red (dark color) and those at the lower edge are shown in
green (light color).  Although the results are shown for an armchair-type edge, analogous edge states are formed
for other edge types.\cite{footnotezigzag}

At $\lambda = \lambda_{c}$, the quasi-energy gap closes at
$\epsilon = \pm \pi/T$. Since the gap centered at $\epsilon=0$
remains open at this point, the chiral edge states \emph{cannot}
be removed as $\lambda$ is increased through $\lambda_{c}$.
Remarkably, the system supports topologically-protected chiral
edge states for $\lambda > \lambda_c$, despite the fact that the
Chern numbers of \emph{both} bulk bands are zero.
The chiral edge states in the quasi-energy spectrum for $\lambda = 4 > \lambda_{c}$ are shown in Fig.\ref{chiraledge}d).
Isolating the two edges, note that each edge state has a ``winding'' structure which resembles that shown in Fig. \ref{1ddynamicaltopologymix}a).
This suggests that these edge modes may be characterized in terms of the topological invariant $\nu_1$ studied in Sec. \ref{1dinvariant}.

\begin{figure}[t]
\begin{center}
\includegraphics[width = 5cm]{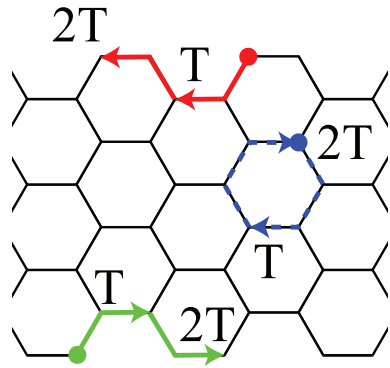}
\caption{Dynamics under the driving cycle shown in Fig.\ref{hexagonallatticedynamics}, in the limit $J \rightarrow 0, \lambda \rightarrow \infty$ with $\lambda J (T/3) = \pi/2$.
The red (green) solid line shows the unidirectional propagation of a particle initially localized on the $B$ ($A$) sublattice along the upper (lower) edge.
As indicated by the blue dotted line, all bulk states are localized.
A particle initially localized at any site in the bulk returns to its original position after a time $2T$. }
\label{nu1edge}
\end{center}
\end{figure}

In order to understand the origin of these features, it is helpful
to consider the limit $J \rightarrow 0$, while keeping $\lambda J
T/3 = \pi/2$ non-zero. In this limit, during each of the three stages
of the cycle, only sites connected by the bold lines in Fig.
\ref{hexagonallatticedynamics} are coupled. The amplitude of
hopping between the two lattice sites of a dimer is chosen such
that after a time period of $T/3$, a particle is transferred with
certainty from one lattice site to its neighbor. The dynamics in
this limit is depicted in Fig. \ref{nu1edge}, where the red and
green solid lines show the propagation of particles along the
upper and lower edges, respectively, and the blue dotted line
shows the propagation of particles in the bulk. Note that all the
Floquet states in the bulk are localized: a particle starting at
any site in the bulk comes back to the starting site after two
complete driving periods. On the other hand, a particle which
starts at the upper (lower) edge on sublattice $B$ ($A$) propagates
unidirectionally along the edge to the left (right). Therefore, the
bulk and edge states are clearly separated.

 In the limit described above, it is straightforward to confirm that the Floquet evolution operators $\{U_{k_{x}}(T)\}$  
restricted to the space spanned by sites on sublattice $B$ ($A$) of the upper (lower) edge, yield $\nu_{1}=1(-1)$.
These chiral Floquet edge states cannot be removed unless the bulk quasi-energy gap closes.
Thus, for any other values of $J$ and $\lambda$ which can be reached from this limit without closing the bulk gap, the existence of chiral Floquet edge states is guaranteed.
This is indeed the case for $\lambda = 4$ and $JT = \pi/16$, whose quasi-energy spectrum is shown in Fig.\ref{chiraledge}.

 Recently, there have been several proposals to induce topological phases with AC electromagnetic fields\cite{oka, Lindner,tanaka}.
In spirit, these models are similar to ours in the small ($\lambda - 1$) limit, where the bulk bands possess non-trivial topology (in our case and in Ref.\onlinecite{oka} characterized by the first Chern number, and in Refs.\onlinecite{Lindner,tanaka} characterized by a $Z_2$ invariant).
In particular, all of these models support robust chiral or helical Floquet edge states.
However, the existence of chiral edge modes {\it without} non-trivial topological bulk quasi-energy bands is a unique feature of our model, which displays a non-trivial winding of quasi-energy.
\section{Discussion} \label{conclusion}

We have presented a unified framework for the topological classification of periodically-driven systems.
Such systems are characterized in terms of their Floquet operators, {\it i.e.} their unitary evolution operators acting over a full period of driving.
Non-trivial topology can arise on two levels.
First, we characterize the topological structure of the Floquet operator in terms of the homotopy groups of unitary matrices.
Here, topologically distinct classes of dynamics are identified by the values of (generalized) ``winding numbers,'' $\nu_1$ and $\nu_3$ for the first and third homotopy groups.
Explicit expressions for $\nu_1$ and $\nu_3$ are given
for systems with discrete
translational symmetry in Eqs.(\ref{nu1}) and (\ref{nu3}), and in
terms of a more general twisted boundary condition formulation in
Eqs.(\ref{gnu1}) and (\ref{gnu3}). Within this framework, the
well-known phenomenon of quantized charge transport\cite{thouless}
obtains a natural and intuitive explanation in terms of the
winding of quasi-energy. Systems with topologically non-trivial
Floquet operators in this homotopy sense can exhibit interesting
features, such as a chiral dispersion in one-dimension, or
protected edge states in two-dimensional systems with
topologically trivial bulk band structures. Such phenomena cannot
be realized in systems governed by local, static
(time-independent) Hamiltonians.

When the Floquet operator is ``trivial'' in this homotopy group sense, 
then the long-time dynamics of the system can be described in terms of a local, time-independent effective Hamiltonian.
In this case, the standard classification schemes of time-independent topological phases can be
applied\cite{kitaev,schnyder,qi}.
Therefore, analogous phases to those of topological
insulators and superconductors can be found in dynamically driven
systems. 
Proposals such as those of Refs.\onlinecite{oka, tanaka,Lindner}, in which
topologically non-trivial states are induced dynamically in otherwise trivial systems, belong to this class.

This work opens many avenues for future exploration.
In particular, it will be interesting to search for experimental manifestations of quasi-energy winding in artificial and natural systems.
Using quantum walks, it is already possible to realize topological phases in one dimension\cite{kitagawa}, and we expect that more complex examples will soon be accessible.
Additionally, the prospect of dynamically inducing topological states in naturally occuring, topologically trivial, materials is particularly intriguing.

In this paper, we have focused on single-particle dynamics in non-interacting systems.
As shown in the appendices, the results can also be applied to the adiabatic dynamics of gapped many-body systems.
It is interesting to extend this work to more general many-body systems, and to open systems subject to decoherence.
These extensions could pave the way to finding new methods of robust quantum control in many-body systems.

\acknowledgements{We thank Bertrand Halperin, Michael Levin, Hosho
Katsura, Liang Fu, Netanel Lindner, Gil Refael and Cenke Xu for
stimulating discussions. This work is supported by 
NSF grant DMR-07-05472, DMR-0757145 (EB),
DMR-090647 and PHY-0646094 (MR), AFOSR Quantum Simulation MURI, AFOSR MURI on
Ultracold Molecules, DARPA OLE program and Harvard-MIT CUA.

\appendix
\section{Topological classification in the presence of disorder} \label{generalizednu1}
Topological properties of physical systems are expected to be
robust against a broad range of perturbations. In sections
\ref{1dinvariant} and \ref{3dinvariant}, we studied the
topological properties of evolution operators 
characterized in terms of homotopy groups.
Non-trivial topology in this homotopy group sense is associated with the ``winding'' of quasi-energy in the Brillouin zone.
In all the cases considered so far, the topological invariants
were defined under the conditions of (discrete) translational
invariance. In particular, Eq.(\ref{nu1}) and Eq.(\ref{nu3}) are
written in terms of the conserved crystal momentum which is
associated with this symmetry.
Below, we generalize the invariant $\nu_1$ to non-translationally
invariant situations, using the idea of twisted boundary
conditions\cite{niu}.
The value of the generalized invariant $\tilde{\nu}_{1}$, defined in
Eq.(\ref{gnu1}) below, gives the same value as $\nu_1$, Eq.(\ref{nu1}), in the
absence of disorder.
An analogous generalization of the invariant $\nu_3$ to the disordered case
can be achieved in a similar manner. 

In the following, we consider a weakly-disordered one dimensional system of a finite length $La$, where $a$ is the lattice constant of the ``unperturbed'' translationally-invariant system in the absence of disorder.
Disorder is weak in the sense that the energy eigenvalues of the instantaneous Hamiltonian $H(t)$ at each time $t$ remain separated into bands which can be associated with the bands of the unperturbed Hamiltonian.
We assume, as before, that the evolution over one full period $T$ only mixes a finite number, $m$, of these bands of the initial Hamiltonian, $H(0)$,
and seek to characterize the topological properties of the finite dimensional $mL\times mL$ evolution operator $U(T)$. 

To probe the topological properties of $U(T)$, we impose ``twisted boundary conditions'' parameterized by an angle $\theta$, such that all wavefunctions must satisfy
$\bra{x=0} \psi \rangle = e^{i\theta} \bra{x=L} \psi \rangle$.
In principle, for each value of $\theta$, one can find the corresponding twisted boundary condition eigenstates of the Floquet operator $U(T)$, and their associated quasi-energies.
However, the inconvenience of dealing with twisted boundary conditions can be eliminated by studying the transformed Floquet operator
\begin{equation}
 U_\theta(T) = e^{-i\hat{x} \theta/L} U(T)e^{i\hat{x} \theta/L},
 \end{equation}
where $U_\theta(T)$ acts on states with {\it periodic} boundary conditions.

In terms of the twist angle $\theta$, we define the generalized topological invariant $\tilde{\nu}_1$ as
\begin{equation}
\tilde{\nu}_{1}  = \frac{1}{2\pi} \int^{\pi}_{-\pi} d\theta\,
\textrm{Tr} \left[ U_\theta(T)^{-1} i\partial_{\theta} U_\theta(T)
\right],  \label{gnu1}
\end{equation}
where the trace is taken over all states in the $mL$-dimensional low-energy subspace (with periodic boundary conditions).
Notice that expression (\ref{gnu1}) for $\tilde{\nu}_{1}$ is analogous to that of $\nu_1$, Eq.(\ref{nu1}), with the phase twist $\theta/L$ playing the role of the crystal momentum.
In this way, the entire disordered system of length $L$ plays the role of one giant unit cell of a much larger, periodic system with $mL$ bands.
As above, $\tilde{\nu}_1$ counts the net winding of quasi-energy as $\theta$ is taken from 0 to $2\pi$ (see Fig.\ref{fig:generalizednu1}).

\begin{figure}[t]
\begin{center}
\includegraphics[width = 5cm]{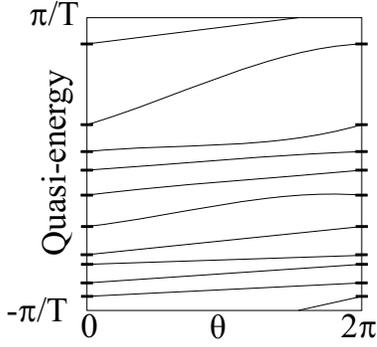}
\caption{The generalized topological invariant $\tilde{\nu}_1$, see Eq. (\ref{gnu1}), counts the net winding
of quasi-energy in one dimension as the twist-angle $\theta$ is taken from $0$ to $2\pi$.
Here, for each $\theta$, we plot the $mL$ quasi-energy eigenvalues of the evolution operator.
}
\label{fig:generalizednu1}
\end{center}
\end{figure}

Similarly, the analogous generalization of $\nu_3$ for a finite system in three dimensions is given by
\begin{eqnarray}
\tilde{\nu}_{3} & = & \iiint  \frac{d^3 \theta}{24 \pi^2}\ \epsilon^{\alpha \beta \gamma}\times \\ &&\textrm{Tr}\Big[  \left(U^{-1}_{\bm{\theta}}\partial_{\theta_\alpha}U_{\bm{\theta}}\right) 
\left(U^{-1}_{\bm{\theta}}\partial_{\theta_\beta}U_{\bm{\theta}}\right) \left(U^{-1}_{\bm{\theta}}\partial_{\theta_\gamma}U_{\bm{\theta}}\right) \Big],\nonumber
\label{gnu3}
\end{eqnarray}
where $\alpha, \beta, \gamma$ label the Cartesian directions $\{x, y, z\}$, $\bm{\theta} = (\theta_{x}, \theta_{y}, \theta_{z})$ is a vector of twist angles for the boundary conditions in each of the three spatial directions, and the integrals are taken over $-\pi \le \theta_\alpha < \pi$.

The generalized topological invariants $\tilde{\nu}_{1}$ and
$\tilde{\nu}_{3}$, defined in terms of twisted boundary conditions, can also be applied to many-body systems, as
long as there is a finite gap between the ground state and the rest of the spectrum throughout the driving period.
In particular, it can be applied in the case of band insulators.
In these cases, $\tilde{\nu}_{1}$ and $\tilde{\nu}_{3}$ are
meaningful even in the presence of many-body interactions\cite{niu}.


\section{Relation between quantized charge pumping and quasi-energy winding} \label{pump}
In this Appendix, we study the topological structure of quantized
adiabatic pumping, first proposed by Thouless\cite{thouless}, as
described in terms of the winding number $\nu_1$. Consider
spinless fermions moving in a periodically varying,
one-dimensional lattice potential. We assume that, at any time
$t$, the $m$ lowest bands of the instantaneous Hamiltonian $H(t)$
are filled, i.e. the system forms a band insulator.
This is the case if the rate of change of the potential is slow
enough such that the evolution is adiabatic with respect to the
band gap at all times, and no particles-hole excitations created.
Thouless proved that, under these circumstances, the number of
particles transported through the system over one driving period
is quantized as an integer\cite{thouless}. Here, we show that this
quantized transport can be naturally understood as a consequence
of the winding of quasi-energies of the corresponding evolution
operator, as captured by the $\nu_1$ invariant.
In fact, such quantized charge transport amounts to a 
realization of the quantized average velocity (or displacement) discussed 
in Section \ref{1dinvariant}. 



We begin by relating $\nu_1$ to the charge current integrated over
one period of the dynamics.
Assuming that the particles do not interact, we study the properties of the single-particle dynamics.
For simplicity we assume translational invariance, but for interacting or disordered systems, pumping of the many-body state can be studied using the twisted boundary condition method described in Appendix \ref{generalizednu1}.

It is useful to work in the basis of cell periodic states
$\ket{u_k, \alpha} = e^{-ik\hat{x}}\ket{k, \alpha} $, where
$\ket{k, \alpha}$ is the Bloch wave function with crystal momentum
$k$ and band index $\alpha$. The corresponding transformed
Hamiltonian is given by $\tilde{H}_k(t) = e^{-ik\hat{x}} H(t)
e^{ik\hat{x}}$, and the evolution operator is given by
$\tilde{U}_k(t) =e^{-ik\hat{x}} U(t) e^{ik\hat{x}}$. The Floquet
evolution operator $\tilde{U}_k(T)$, which acts on the wave
functions $\{\ket{u_k, \alpha}\}$ with periodic boundary
conditions, only mixes the filled bands due to the adiabaticity
condition of the dynamics. Therefore, when restricted to the
manifold of filled bands, and projected onto the subspace with a
particular crystal momentum $k$, $\tilde{U}_k(T)$ reduces to
$U_{k}(T)$ as defined in the main text.

The instantaneous pumped current in the many-body insulating state can be expressed as
\begin{eqnarray*}
J(t) &=&  \sum_{\alpha \in \mathrm{occ.}} \int^{\pi}_{-\pi} \frac{dk}{2\pi}   \bra{u_k, \alpha} \frac{\partial \hat{x}(t)}{\partial t}  \ket{u_k, \alpha} \\
&=& \sum_{\alpha \in \mathrm{occ.}} \int \frac{dk}{2\pi}
\bra{u_k, \alpha}
\tilde{U}_k(t)^{-1} i [\tilde{H}_k(t), \hat{x}] \tilde{U}_k(t)\ket{u_k, \alpha} \\
&=& \sum_{\alpha \in \mathrm{occ.}} \int \frac{dk}{2\pi}
\bra{u_k, \alpha} \tilde{U}_k(t)^{-1} \left( \partial_k
\tilde{H}_k(t) \right) \tilde{U}_k(t)\ket{u_k, \alpha},
\end{eqnarray*}
where we have used the Heisenberg representation, and the
summation of $\alpha$ is over the occupied bands. Using the
relation $\partial_k \tilde{U}_k(T) = - \int^{T}_{0}
\tilde{U}_k(T-t) \left(i \partial_k  \tilde{H}_k(t) \right)
\tilde{U}_k(t) dt$, the total pumped current integrated over one period is
\begin{eqnarray}
\label{Jtot}
\int^{T}_{0} dt &&\!\!\!\!\!\!\!J(t) =\\ && \!\!\!\! \sum_{\alpha \in \mathrm{occ.}} \int^{\pi}_{-\pi} \frac{dk}{2\pi}
  \bra{u_k, \alpha} \tilde{U}_k(T)^{-1} \left(i\partial_k
\tilde{U}_k(T) \right) \ket{u_k, \alpha}. \nonumber
\end{eqnarray}
Equation (\ref{Jtot}) is equivalent to the expression of $\nu_1$ in
Eq.(\ref{nu1}), with the understanding that \emph{the trace is
taken only over the occupied bands.}
As pointed out in Sec. \ref{1dinvariant}, $\nu_1$ counts the
winding number of the quasi-energy bands; hence the total pumped charge can
take only integer values.

To demonstrate the quasi-energy winding of a system that displays
quantized pumping, we consider the following example:
\begin{equation}
H(t) = \frac{\hat{p}^2}{2m} + \lambda \cos(2\pi \hat{x}-\omega t),
\end{equation}
where we have taken the lattice constant to be $1$.
The cosine potential is used for concreteness, although the
following argument applies to any moving periodic potential
$V(2\pi \hat{x}-\omega t)$. The (real) momentum is denoted by $p$,
which should not be confused with crystal momentum $k$.

The evolution operator under the time-dependent Hamiltonian above
can be exactly obtained by a Galilean transformation to the comoving
frame. It is straightforward to check that the following evolution
operator satisfies the equation of motion $i \partial_t U(t) =
H(t) U(t)$;
\begin{equation}
U(t) = e^{-i\omega t \hat{p}/(2\pi)} e^{-i H_{0} t}.
\end{equation}
Here, $H_{0}$ is the static Hamiltonian $H_0 = \frac{p^2}{2m} -
\frac{\omega \hat{p}}{2\pi}+ \lambda \cos(2\pi x)$. 
The evolution operator after one period can be written as
\begin{equation}
U(T) = e^{-i \hat{p}} e^{-i H_{0} T},
\end{equation}
where $T = 2\pi/\omega$. Let $\{\ket{k,\alpha}\}$ be the Bloch
eigenstates for $H_{0}$, where $k$ is the crystal momentum and
$\alpha$ is the band index. 
By definition, $\ket{k,\alpha}$ is an eigenstate of the shift
operator $e^{-i \hat{p}} $ with eigenvalue $e^{-i k}$. Therefore,
we may write
\begin{equation}
U_{k}(T)\ket{k,\alpha} = e^{-i k} e^{-i E_{k,\alpha} T}\ket{k,\alpha},
\end{equation}
where $E_{k,\alpha}$ is the energy of the Bloch state
$\ket{k,\alpha}$. Since $E_{k,\alpha}$ is a periodic function of
$k$, it is clear from this expression that each quasi-energy band
of $U(T)$ is characterized by a winding number of 1.

Note that in the example above, the adiabaticity condition was not
used -- the quasi-energy winding is a straightforward consequence
of the Galileian invariance of the system. Even if Galileian
invariance-breaking perturbations, such as a small
time-independent potential, are added, as long as the adiabaticity
condition is imposed ($\omega \ll \Delta E_g$, where $\Delta E_g$
is a typical band gap), particles cannot be excited from one band
to the other during the dynamics. Therefore, the quasi-energy
winding of each band cannot change.



\section{Relation between the topological invariants of evolution operators and
Chern numbers} \label{chernnumbers} The Chern numbers have been
used intensively to classify insulators and superconductors. The
first Chern number is defined in two dimensions, and was used to
explain the quantization of the Hall coefficients in integer
quantum Hall systems\cite{TKNN}. More recently, Qi \emph{et.
al.}\cite{qi} have shown that topological invariants of two and
three dimensional topological insulators are closely related to
the second Chern number in four dimensions.

In this paper, we have shown that periodically driven systems can be
characterized by the invariants $\nu_1$ and $\nu_3$, which are
written entirely in terms of evolution operators.
Through dimensional reduction\cite{qi}, we establish a connection between $\nu_1$ and
$\nu_3$, and the first and second Chern numbers in $2$ and $4$
dimensions, respectively.

Qi \emph{et. al.}\cite{qi} classified a family of $2n-1$
dimensional insulators indexed by a compact parameter $\theta$ through Chern
numbers defined in $2n$ dimensions by regarding the parameter
$\theta$ as an additional component of crystal momentum.
In this way, they have shown
that the first and second Chern numbers defined for a family of
$1$ and $3$
dimensional insulators are given by
\begin{eqnarray}
C_{2} & = & \frac{1}{2\pi} \oint d\theta  \frac{\partial P_{1}(\theta)}{\partial \theta} \label{firstchern} \\
C_{4} & = &  \frac{1}{16\pi^2} \oint d\theta  \frac{\partial
P_{3}(\theta)}{\partial \theta}, \label{secondchern}
\end{eqnarray}
where
\begin{eqnarray}
P_{1}(\theta) & = &  \int dk  \textrm{Tr} \left[ a_{x}(\theta, k)  \right] \label{polarization1} \\
P_{3}(\theta) & = &
 \int d^3 \vec{k} \epsilon^{ijk}   \textrm{Tr} \left[  \left( f_{ij} - \frac{i}{3} [a_i, a_j] \right) \cdot a_k \right],  \nonumber\\
&& \label{polarization3} \\
\end{eqnarray}
and
\begin{eqnarray}
a_{i}^{\alpha \beta}(\theta, \vec{k})& =&  (-i) \bra{u_{\vec{k}}, \theta, \alpha} \partial_{k_i} \ket{u_{\vec{k}}, \theta, \beta} \label{berryphase}\\
f_{ij}^{\alpha \beta}(\theta, \vec{k}) &=& \partial_{k_i}
a_{j}^{\alpha \beta} - \partial_{k_j} a_{i}^{\alpha \beta} +
i[a_i, a_j]^{\alpha \beta},
\end{eqnarray}
where  $\epsilon^{ijk}$ is the anti-symmetric tensor and the $\vec{k}$
integration extends over the first Brillouin zone.
Here $\{\ket{u_{\vec{k}}, \theta, \alpha}\}$ is the set of filled states of the system, which satisfy $\ket{u_{\vec{k + G}},\theta,\alpha} = \ket{u_{\vec{k}}, \theta, \alpha}$, where $\vec{G}$ is a reciprocal lattice vector.
Note that the Berry connection $a_{i}$ is a matrix indexed by the internal state labels,
$\alpha$ and $\beta$. The trace is taken over the filled bands.

For \emph{periodically driven} systems, we interpret the
parametric dependence of Bloch states on $\theta$ as coming from
the evolution of the states under the periodic Hamiltonian. Then,
it is possible to write the Chern numbers in Eq.(\ref{firstchern})
and Eq.(\ref{secondchern}) entirely in terms of the evolution
operator of Bloch wave functions after one period, $U_{k}(T)$. We
identify the cyclic parameter $\theta$ with the time variable $t$,
and write the Chern numbers as
\begin{eqnarray}
C_{2} & = & \frac{1}{2\pi} \left( P_{1}(T) - P_{1}(0) \right) \label{c2} \\
C_{4} & = &  \frac{1}{16\pi^2}\left( P_{3}(T) - P_{3}(0) \right)  \label{c4}
\end{eqnarray}
Therefore, the Chern numbers have the meaning of the change of the
``polarizations'' $P_{1}$ and $P_{3}$ after one period.

The polarizations $P_{1}(T)$ and $P_{3}(T)$ at time $t = T$ are related to $P_{1}(0)$ and $P_{3}(0)$ at time
$t=0$ through the evolution of the Bloch states. 
If the Hamiltonian changes slowly compared to the band gap, it
does not mix the occupied states with states above the gap.
Each occupied state $\ket{u_{\vec{k}},0,\alpha}$ at time $t = 0$ evolves into a typically different occupied state $\ket{u_{\vec{k}},T,\alpha}$ at time $t = T$, given by
 $\ket{u_{\vec{k}}, T, \alpha} = U_{\vec{k}}(T)\ket{u_{\vec{k}},0,\alpha} = \sum_{\beta} U_{\vec{k}}^{\beta\alpha}(T) \ket{u_{\vec{k}}, 0, \beta}$,
where $U_{\vec{k}}(T)$ is the evolution operator for crystal momentum $\vec{k}$ and $U_{\vec{k}}^{\beta\alpha}(T) = \bra{u_{\vec{k}},0,\beta}U_{\vec{k}}(T)\ket{u_{\vec{k}},0,\alpha}$.
The evolution of the Bloch states is characterized by a non-abelian Berry
connection $a_{i}^{\alpha \beta}(T)$, see Eq.(\ref{berryphase}) with $\theta$ replaced by $T$: $$a_{i}^{\alpha \beta}(T) =a_{i}^{\alpha \beta}(0) - i
\bra{u_{\vec{k}},0,\alpha} U_{\vec{k}}(T)^{-1}
\partial_{k_{i}}U_{\vec{k}}(T) \ket{u_{\vec{k}},0,\beta}.$$
Substituting
this relation into the definitions of the two polarizations, it is now
straightforward to show that the Chern numbers in Eq.(\ref{c2})
and Eq.(\ref{c4}) are expressed in terms of evolution operators
$U_{\vec{k}}(T)$ and give the topological invariants of evolution
operators $\nu_1$ and $\nu_3$ in Eq.(\ref{nu1}) and
Eq.(\ref{nu3}).

This construction shows the connections between driven dynamics in
$2n-1$ dimensions and $n$th Chern numbers for $n=1,2$. We
emphasize that the dynamics is not necessarily adiabatic with
respect to the separation between the low-lying bands. The only
condition of the applicability of $\nu_1$,$\nu_3$ is that the
dynamics is adiabatic with respect to a band gap which separates
the low-lying bands from all higher states (Fig. \ref{bandgap}),
and therefore only the low-lying bands are mixed after one period.


\section{Time reversal symmetry in periodically driven systems}
\label{sec:TR} In this Appendix, we prove that if the
time-dependent Hamiltonian $H(t)$ of a periodically driven system
satisfies
\begin{equation}
\mathcal{\tilde Q}H(t+t_0)^* \mathcal{\tilde Q}^\dagger =
H(-t+t_0), \label{HTR}
\end{equation}
where $\mathcal{\tilde Q}$ is a unitary operator and $t_0$ is
an arbitrary time, then the evolution operator of the system over
one period is time reversal symmetric [in the sense of Eq.(\ref{TR})] and so is the effective Hamiltonian, Eq.(\ref{heff}). To
show this, we write the evolution operator in the form
\begin{equation}
U(T,0)=\lim_{N \rightarrow \infty} e^{-i\Delta t H(T)} e^{-i\Delta
t H(T-\Delta t)} \dots e^{-i\Delta t H(0)}, \label{UT0}
\end{equation}
where $U(t_1,t_2)$ is the evolution operator from time $t_1$ to
$t_2$, and $\Delta t = T/N$. Using Eq.(\ref{HTR}), and the fact
that $H(t+T)=H(t)$, one can show that
\begin{equation}
\mathcal{\tilde Q} U(T,0)^* \mathcal{\tilde Q}^\dagger =
U(2t_0+T, 2t_0)^\dagger, \label{UT1}
\end{equation}
Then, defining $\mathcal{Q}\equiv U(0,2t_0)
\mathcal{\tilde{Q}}$, we get that
\begin{equation}
\mathcal{Q} U(T,0)^* \mathcal{Q}^\dagger = U(T, 0)^\dagger,
\label{UT2}
\end{equation}
which is Eq.(\ref{TR}).


\begin{thebibliography}{10}
\bibitem{footnote_basis}
Note that evolution operators are only defined up to the choice of the basis
  and two unitary matrices $U_{k}$ and $U'_{k} = V^{-1}_{k} U_{k} V_{k}$
  ($V_{k}$ is a unitary matrix) that are related by $k$ dependent similarity
  transformations represent the same evolution. It is straightforward to show
  that $\nu_{1}$ is invariant under such similarity transformations, and
  therefore, is independent of the choice of basis.
\bibitem{degeneracy}
When there are degeneracies in the spectrum, one should choose the definitions
  of $\{\epsilon_{k,\alpha}\}$ such that they become smooth as a function of
  quasi-momentum $k$.
\bibitem{footnote_homotopy}
If the quasi-energy spectrum is gapped, it is always possible to continuously
  deform the spectrum so that all of the bands become flat (i.e. independent of
  the crystal-momentum) and degenerate. After this deformation, the evolution
  operator becomes proportional to unity for all $\mathbf{k}$. Such an
  evolution operator is clearly topologically trivial. On the other hand, if
  the quasi-energy spectrum is gapless, this procedure cannot be carried out
  because the spectrum has to be periodic in $\mathbf{k}$ throughout the
  deformation.
\bibitem{footnoteeffectiveh}
More general Hamiltonians with two bands can have a sublattice-independent
  energy shift, but determinant of evolution operators resulting from the
  dynamics considered in this section is unity, and therefore, this term is
  absent here.
\bibitem{footnotezigzag}
A similar calculation can be also carried out with arm-chair edges. Dynamical
  induction of Chern number as well as energy winding $\nu_1$ also occur for
  such edges as well.
\end{thebibliography}
\end{document}